\documentclass[conference]{IEEEtran}
\IEEEoverridecommandlockouts
\usepackage{cite}
\usepackage{amsmath,amssymb,amsfonts}
\usepackage{algorithmic}
\usepackage{bm}
\usepackage{graphicx}
\usepackage{pgfplots}
\usepgfplotslibrary{groupplots}
\usepackage{pgfplotstable}
\usepackage{subfig}
\usepackage[mode=buildnew]{standalone}
\usepackage{textcomp}
\usepackage{tikz}
\usepackage{url}
\usepackage{xcolor}
\def\BibTeX{{\rm B\kern-.05em{\sc i\kern-.025em b}\kern-.08em
    T\kern-.1667em\lower.7ex\hbox{E}\kern-.125emX}}

\DeclareMathOperator*{\grad}{\bm \nabla\!}

\begin{document}

\title{JPEG Meets PDE-based Image Compression
\thanks{This work has received funding from the European Research Council 
	(ERC) under the European Union's Horizon 2020 research and 
	innovation programme (grant agreement no. 741215, ERC Advanced 
	Grant INCOVID).}
}

\author{\IEEEauthorblockN{Sarah Andris, Joachim Weickert, Tobias Alt, and Pascal Peter}
\IEEEauthorblockA{\textit{Mathematical Image Analysis Group} \\
\textit{Faculty of Mathematics and Computer Science} \\
\textit{Saarland University}\\
Campus E1.7, 66041 Saarbr\"ucken, Germany \\
\{andris, weickert, alt, peter\}@mia.uni-saarland.de}
}


\maketitle

\begin{abstract}
Inpainting-based image compression is emerging as a promising competitor to 
transform-based compression techniques. Its key idea is to reconstruct image 
information from only few known regions through inpainting. Specific partial 
differential equations (PDEs) such as edge-enhancing diffusion (EED) give high 
quality reconstructions of image structures with low or medium texture. Even 
though the strengths of PDE- and transform-based compression are complementary, 
they have rarely been combined within a hybrid codec. We propose to sparsify 
blocks of a JPEG compressed image and reconstruct them with EED inpainting. Our 
codec consistently outperforms JPEG and gives useful indications for 
successfully developing hybrid codecs further. Furthermore, our method is the 
first to choose regions rather than pixels as known data for PDE-based 
compression. It also gives novel insights into the importance of corner regions 
for EED-based codecs.
\end{abstract}


\section{Introduction}
With image data steadily becoming more abundant and of higher resolution, the  
need for good lossy compression codecs increases. Established transform-based 
codecs such as \linebreak JPEG \cite{PM92} and JPEG2000 \cite{TM02} enforce sparsity in 
the transform domain to reduce coding costs. 
In contrast, inpainting-based codecs \cite{GWWB08} compress data in the spatial domain by carefully selecting and storing only very few pixels. The missing data are reconstructed via inpainting from these pixels, the so-called inpainting mask. Picking a suitable inpainting operator as well as an efficient method for storing the locations of the remaining pixels is vital for the efficiency of a codec.
Diffusion processes modelled by partial differential equations (PDEs) have been particularly successful. The current state of the art for natural colour images is the R-EED-LP codec by \linebreak Peter et al.~\cite{PKW17} which employs edge-enhancing diffusion EED~\cite{We97}. It extends the work of Gali\'{c} et al.~\cite{GWWB08} and \linebreak Schmaltz et al.~\cite{SPME14} and consistently outperforms JPEG. It can also surpass JPEG2000 for high compression ratios and images with medium amounts of texture.

The strengths of transform- and PDE-based codecs are complementary: Diffusion-based inpainting benefits from the inherent smoothness conditions and edge preservation properties of the inpainting operator. Therefore, it can compress piecewise smooth structures highly efficiently by only storing very few mask points \cite{PW15}. However, these codecs are less efficient for high-frequent, repeating structures such as textures. In such regions, transform-based methods in general surpass PDE-based techniques.

\subsection{Our Contribution}

In the present paper, we propose to combine transform-based with PDE-based 
methods to exploit their individual strengths. To this end, we successively 
remove blocks from a JPEG compressed image with a probabilistic sparsification 
approach and reconstruct them with EED inpainting.
The only significant storage overhead is a block mask, indicating which compression technique is used for each block. In particular, we do not use any additional information apart from the JPEG compressed blocks for our inpainting process. The resulting \textit{hybrid block-based EED} codec (\textit{B-EED}) consistently outperforms JPEG, showing that transform-based and inpainting-based ideas can be mutually beneficial. Additionally, B-EED is the first codec to combine stored regions instead of single pixels with EED inpainting. The acquired masks automatically select regions containing corners. This indicates that corner information is not only essential for the reconstruction quality of EED, but also a preferred form of information.

\subsection{Related Work}

Hybrid ideas have been successfully used in image compression in various ways. 
Peter and Weickert \cite{PW15} proposed a block-based decomposition of the 
image by choosing for each block between a reconstruction with EED or 
exemplar-based inpainting. In a similar way, Zhou et al.~\cite{ZYCH19} store
some image blocks with vector quantisation, discard all others and recover them 
with total variation inpainting. However, none of these codecs involve 
transform-based ideas.

In transform-based compression, JPEG \cite{PM92} is currently the most widely used image codec. Its core idea is to apply the discrete cosine transform (DCT) to an image and quantise high-frequent coefficients in a fairly coarse way. This allows a compact data representation without perceptually severe degradations. In order to localise the inherently global transform, it performs all computations on separate $8\times8$ pixel blocks. JPEG is a perfect candidate for integration into a hybrid transform-inpainting codec, since it is easily accessible and structures the image into almost independent blocks.

Building onto JPEG compression, Rane et al.~\cite{RSB03} proposed to 
reconstruct missing blocks with either structure inpainting or texture 
synthesis. More recently, Couto et al.~\cite{CNP17} combined JPEG with 
patch-based inpainting. Following a more complex approach, Liu et 
al.~\cite{LSW12} augmented H.264 \cite{SW05} with both patch-based inpainting 
and additionally stored side information.  All of these methods remove blocks 
either by simple mode detection in scanline order or by analysing image 
structures.

In contrast, we include the reconstruction error directly into our block 
removal approach. Moreover, we generalise the probabilistic sparsification and 
nonlocal pixel exchange methods by Mainberger et al.~\cite{MHWT12} to obtain an 
error-based block selection strategy.  

\subsection{Paper Structure}

In Section \ref{sec:overview}, we briefly review EED-based compression. EED plays the central role in our hybrid codec, which we propose in Section \ref{sec:proposed}. After evaluating our codec in several experiments in Section \ref{sec:experiments}, we conclude with Section \ref{sec:conclusions}.

\section{EED in Image Compression}
\label{sec:overview}

Edge-enhancing diffusion for denoising has been introduced by Weickert~\cite{We94e} and has firstly been used for inpainting by Weickert and Welk~\cite{WW06}. Assume that pixel values of a greyscale image are known at mask positions $K \subset \Omega$ as a subset of the image domain $\Omega \subset \mathbb{R}^2$. Weickert and Welk~\cite{WW06} compute the inpainting result as the steady state of the image $u(x,y,t)$ that evolves under the EED equation
\begin{equation}
\partial_t u = \text{div}\left(\bm D(\bm \nabla u_\sigma)\bm \grad u\right) \quad \text{on} 
\quad \Omega \backslash K \times \left[0, \infty\right),
\end{equation}
where $u_\sigma$ is a Gaussian-smoothed version of $u$ with standard deviation $\sigma$. 

For EED, Weickert \cite{We94e} defines the diffusion tensor $\bm D$ via its eigenvectors $\bm v_1 \! \parallel \! \bm \nabla u_\sigma$ and $\bm v_2 \! \perp \! \bm \nabla u_\sigma$ and corresponding eigenvalues $\lambda_1$ and $\lambda_2$. The eigenvalues are set as $\lambda_1 = g(|\bm \nabla u_\sigma|^2)$ and $\lambda_2 = 1$ with the Charbonnier diffusivity $g(s^2) = (1 + \frac{s^2}{\lambda^2})^{-\frac{1}{2}}$ \cite{CBAB94}. By design, EED allows smoothing along edges, but inhibits it across them. The contrast parameter $\lambda$ determines how pronounced the gradient magnitude has to be to indicate an edge. Since EED guides diffusion along image structures, it is able to propagate information over large empty image regions. 

Gali\'c et al.~\cite{GWWB08} exploited the strong reconstructing properties of 
EED inpainting for image compression by storing values of very sparse masks. 
Building on this first codec, Schmaltz et al.~\cite{SPME14} optimised 
performance for grey value images in their R-EED codec. For colour images, 
Peter et al.~\cite{PKW17} extended R-EED to their R-EED-LP codec by enforcing 
sparser masks in the chroma channels and guiding inpainting there through the 
diffusion tensor of the reconstructed luma channel. R-EED-LP can outperform 
JPEG2000 for large compression ratios and images with medium texture. For 
highly textured regions, R-EED-LP becomes inefficient, since it has to store an 
extensive amount of mask points. As a remedy, codecs with nonlocal,
exemplar-based inpainting have been proposed \cite{KBPW18}, but they are 
computationally expensive. Thus, let us now explore a faster hybrid 
strategy.

\section{Marrying JPEG and EED Inpainting}\label{sec:proposed}

\setlength{\fboxsep}{0pt}%
\setlength{\fboxrule}{1pt}%

\begin{figure}[!t]
	\subfloat[Original.]{
		\fcolorbox{gray}{white}{\includegraphics[width=0.14\textwidth]{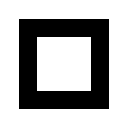}}%
		\label{fig:block-orig}
	}
	\hfil
	\subfloat[Sparsified Image.]{
		\fcolorbox{gray}{white}{\includegraphics[width=0.14\textwidth]{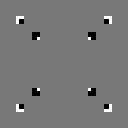}}%
		\label{fig:block-init}
	}
	\hfil
	\subfloat[Reconstruction.]{
		\fcolorbox{gray}{white}{\includegraphics[width=0.14\textwidth]{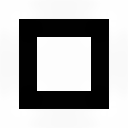}}%
		\label{fig:block-recon}
	}
	\caption{Block sparsification on a simple test image. Discarded blocks are 
		set to grey. B-EED removes all blocks except for those 8 which contain a 
		corner, indicating that corner regions are the preferred form of known data 
		for EED-based codecs.}
	\label{fig:toy-sparsify}
\end{figure}

\begin{figure*}
	\begin{tabular}{ccc}
		\includegraphics[width=0.3\linewidth]
		{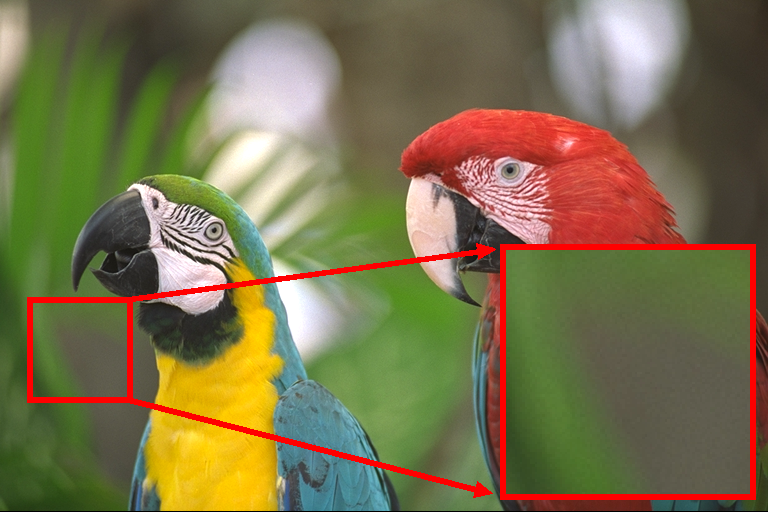}
		&\includegraphics[width=0.3\linewidth]
		{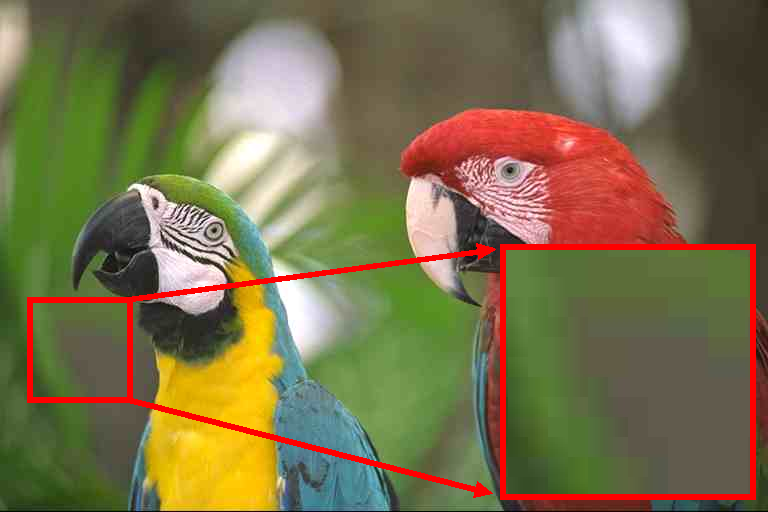}
		&\includegraphics[width=0.3\linewidth]
		{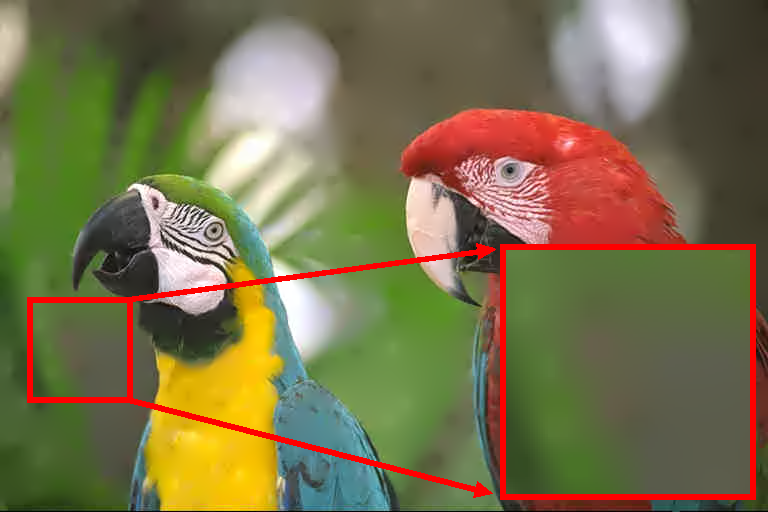} 
		\\
		original image
		& JPEG compressed image 
		& B-EED compressed image 
		\\
		& (ratio 60.6:1, PSNR 33.1 dB) & (ratio 89.6:1, PSNR 32.7 dB) \\[2mm]
		\includegraphics[width=0.3\linewidth]
		{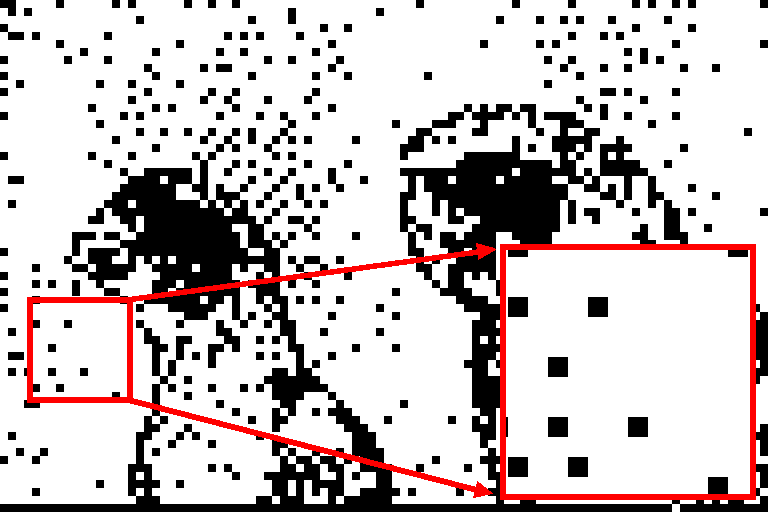}
		&\includegraphics[width=0.3\linewidth]
		{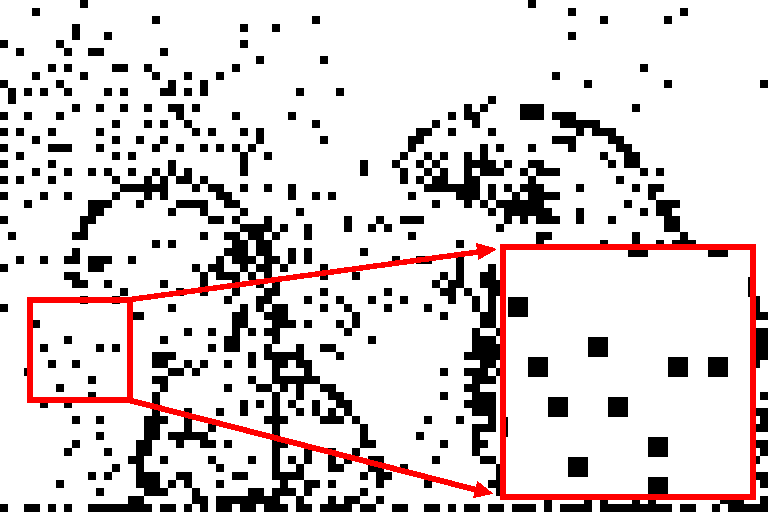}
		&\includegraphics[width=0.3\linewidth]
		{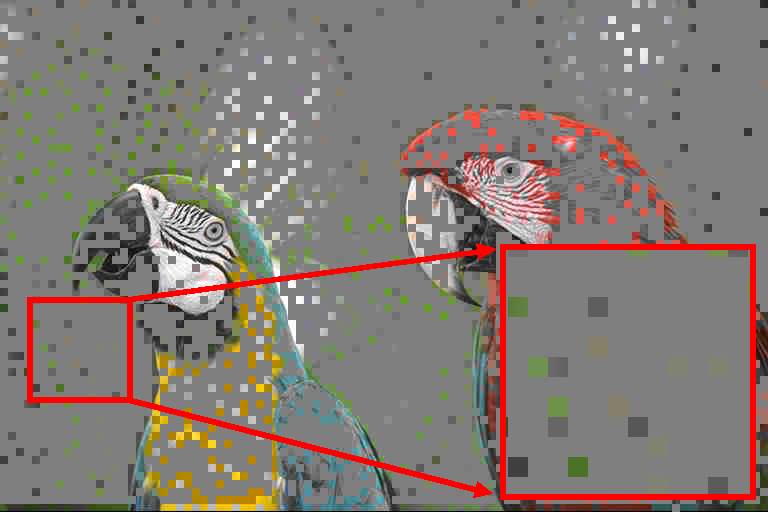}
		\\
		mask for luma channel 
		& mask for chroma channels 
		& inpainting initialisation 
	\end{tabular}
	\caption{Comparison between JPEG and B-EED on a natural image. The bottom right areas are zooms of the corresponding left rectangles. Best viewed in colour. \textbf{Top row:} The B-EED codec reduces coding costs by removing blocks from a baseline JPEG image. 
	\textbf{Bottom row:} Mask blocks (in black) cluster at textured regions. 
		\label{fig:image-comparison}}
\end{figure*}

JPEG performs significantly worse than EED-based codecs on smooth regions since 
it has to store values for blocks where the inpainting process can propagate 
information from surrounding regions. By dropping image data in such areas and 
reconstructing them with EED inpainting, we can reduce storage significantly 
while preserving the overall reconstruction error or even decreasing it. Our 
novel, block-based B-EED codec starts with a pure JPEG compressed version as 
input image. In order to find out which JPEG blocks to remove, we generalise 
the pixel-based probabilistic sparsification of Mainberger et al.~\cite{MHWT12} 
to image blocks.

\subsection{Data Optimisation}

In a probabilistic block sparsification step, we randomly pick a fraction 
$c_\text{PS}$ of all mask blocks as candidates and remove them temporarily from 
the inpainting mask. We then perform one global EED inpainting with the new 
mask. Afterwards, we add back $(1 - r_\text{PS})$ of the removed blocks that 
have the largest local error, as it is probable that these blocks are important 
for the reconstruction. Consequently, we have removed $r_\text{PS}c_\text{PS}$ 
of all mask blocks in one step. We repeat this process until the mask reaches a 
target density. This allows to identify those blocks that we can 
discard with least influence on the quality.

Since the greedy sparsification strategy does not guarantee to reach a globally optimal mask, we additionally introduce nonlocal block exchange (NLBE) as a generalisation of nonlocal pixel exchange (NLPE) \cite{MHWT12}. This allows to reinsert blocks that have been removed in a sparsification step, although they are important for a good reconstruction. First, we randomly pick a fraction $c_\text{NLBE}$ of non-mask blocks as candidates. From these candidates, we pick a fraction $r_\text{NLBE}$ with largest local error and add them back to the mask. In order to keep the mask density constant, we randomly remove $r_\text{NLBE}c_\text{NLBE}$ of blocks again. If the new mask improves the EED reconstruction, we keep it, otherwise we discard it. 

JPEG transforms colour images to YCbCr space and subsamples the chroma channels by a factor 2, exploiting the fact that the human visual system values structure over colour. Since our block sparsification method inherently reduces information in the spatial domain, we can achieve a similar effect by allowing a differing mask density in the chroma channels. B-EED separately optimises the mask density of the luma and the chroma channels w.r.t. reconstruction error. Indeed, the resulting chroma densities are much lower than the corresponding luma densities. 

Moreover, we found that block masks for a fixed density hardly 
change for reasonable choices of the EED parameters $\sigma$ and $\lambda$. 
Thus, we first compute a mask with probabilistic sparsification for standard EED parameters. In order to achieve a given target compression ratio, this step includes the optimisation of the compression ratio of the baseline JPEG image as well as the mask densities of luma and chroma channels. Afterwards, we optimise $\sigma$ and 
$\lambda$ in two nested golden section searches. Subsequently applying NLBE 
eliminates blocks that are not optimal for the new parameter setting. 

\begin{figure*}
	\centering
	\begin{tabular}{c c c}
		Original & JPEG & B-EED \\
		\includegraphics[width=.3\textwidth]{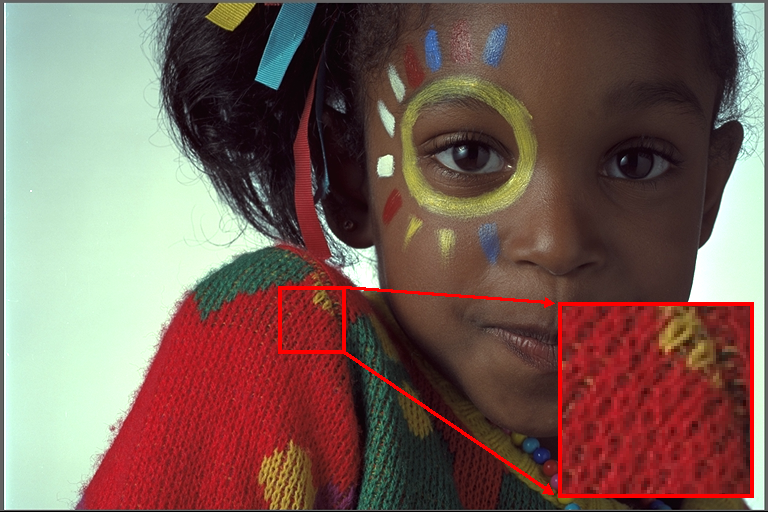} 
		& \includegraphics[width=.3\textwidth]{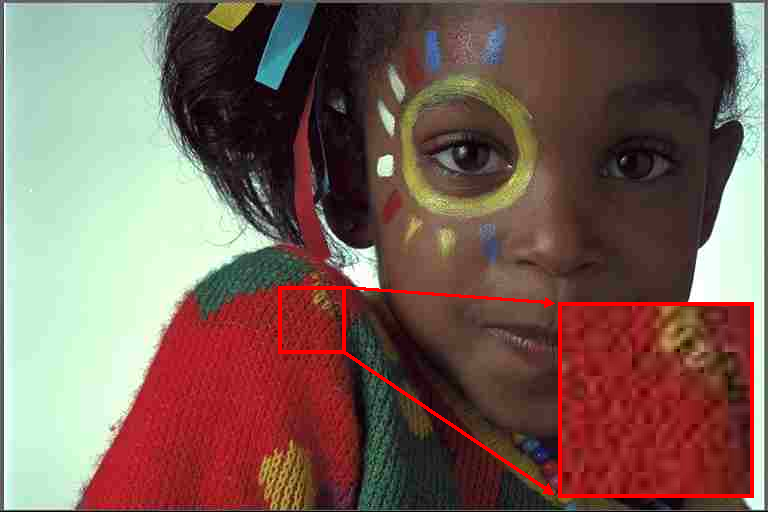}
		& \includegraphics[width=.3\textwidth]{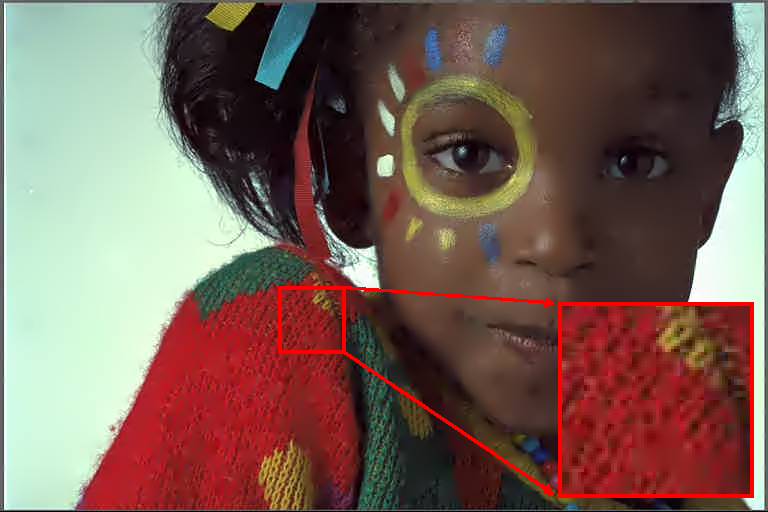} \\
		& ratio 68.4:1, PSNR 29.7 dB & ratio 70.4:1, PSNR 30.9 dB \\[2mm]
		\includegraphics[width=.3\textwidth]{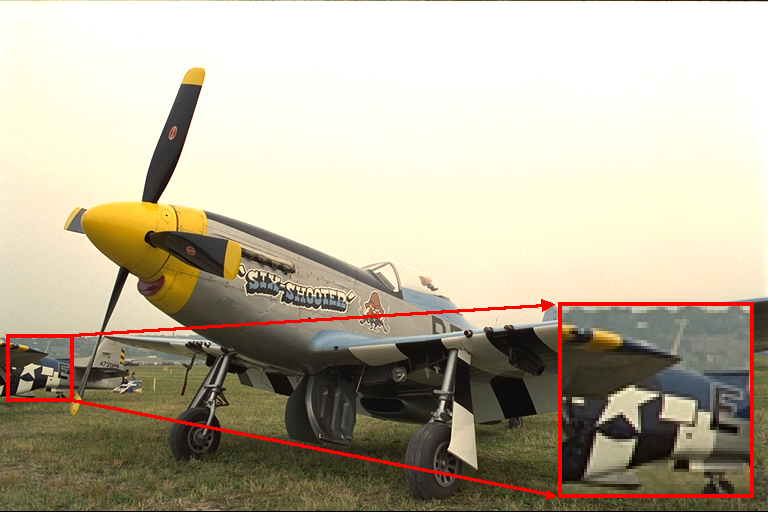} 
		& \includegraphics[width=.3\textwidth]{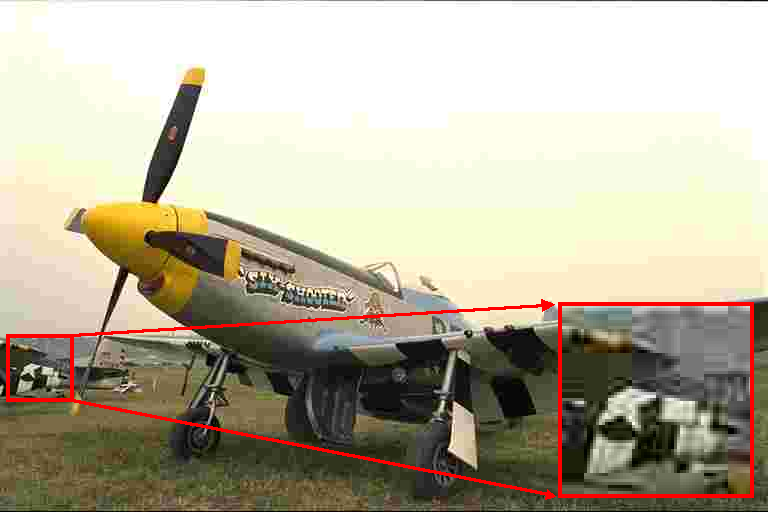}
		& \includegraphics[width=.3\textwidth]{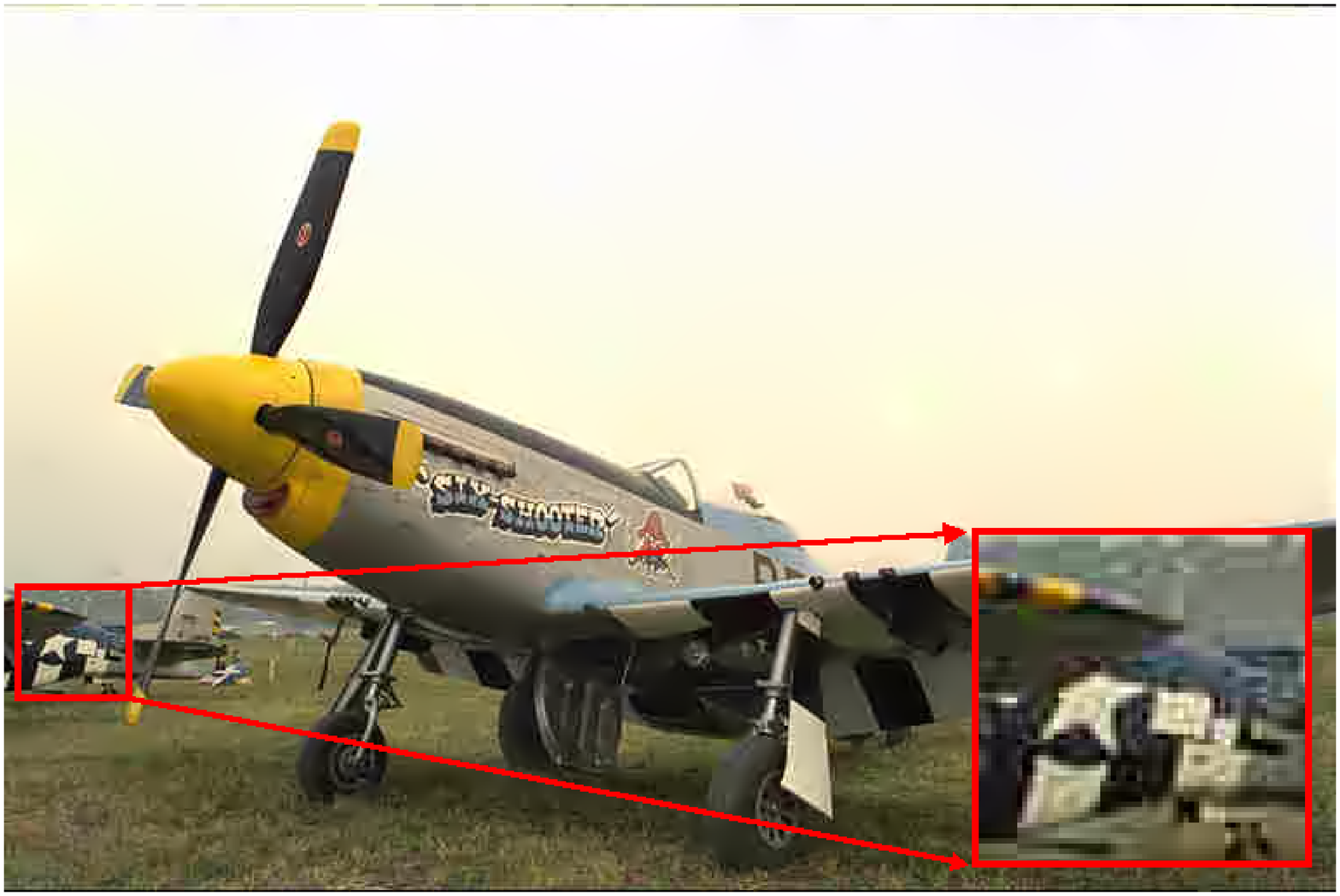} \\
		& ratio 93.1:1, PSNR 28.3 dB & ratio 94.4:1, PSNR 30.0 dB \\[2mm]
		\includegraphics[width=.3\textwidth]{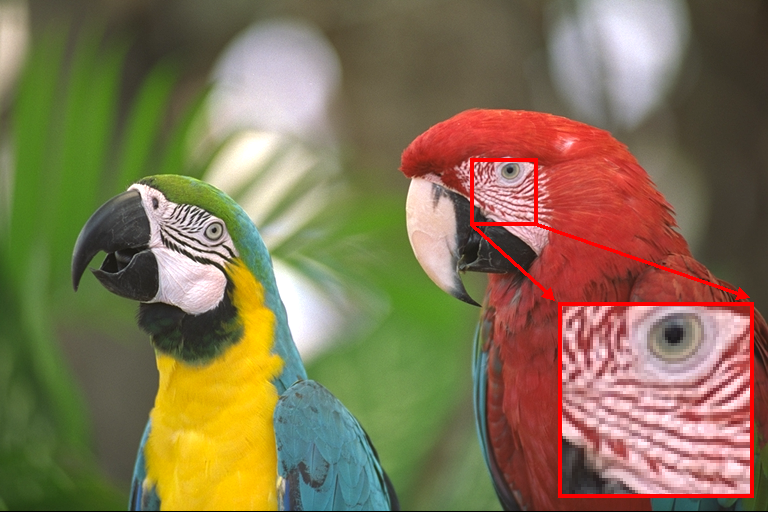} 
		& \includegraphics[width=.3\textwidth]{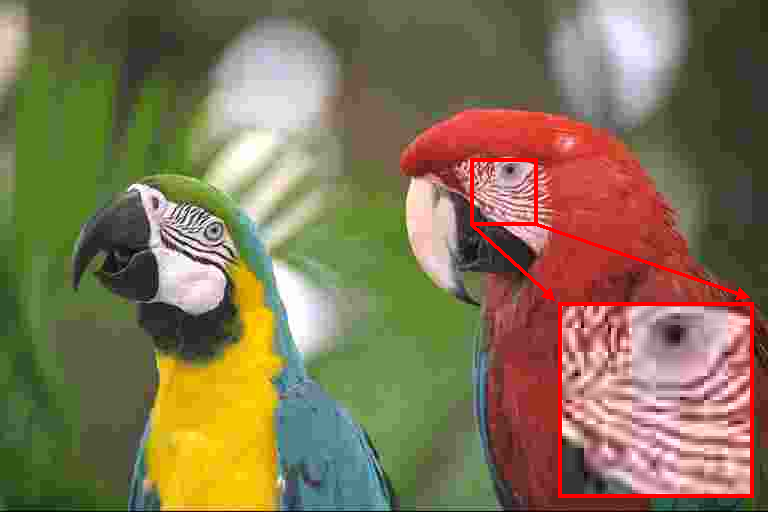}
		& \includegraphics[width=.3\textwidth]{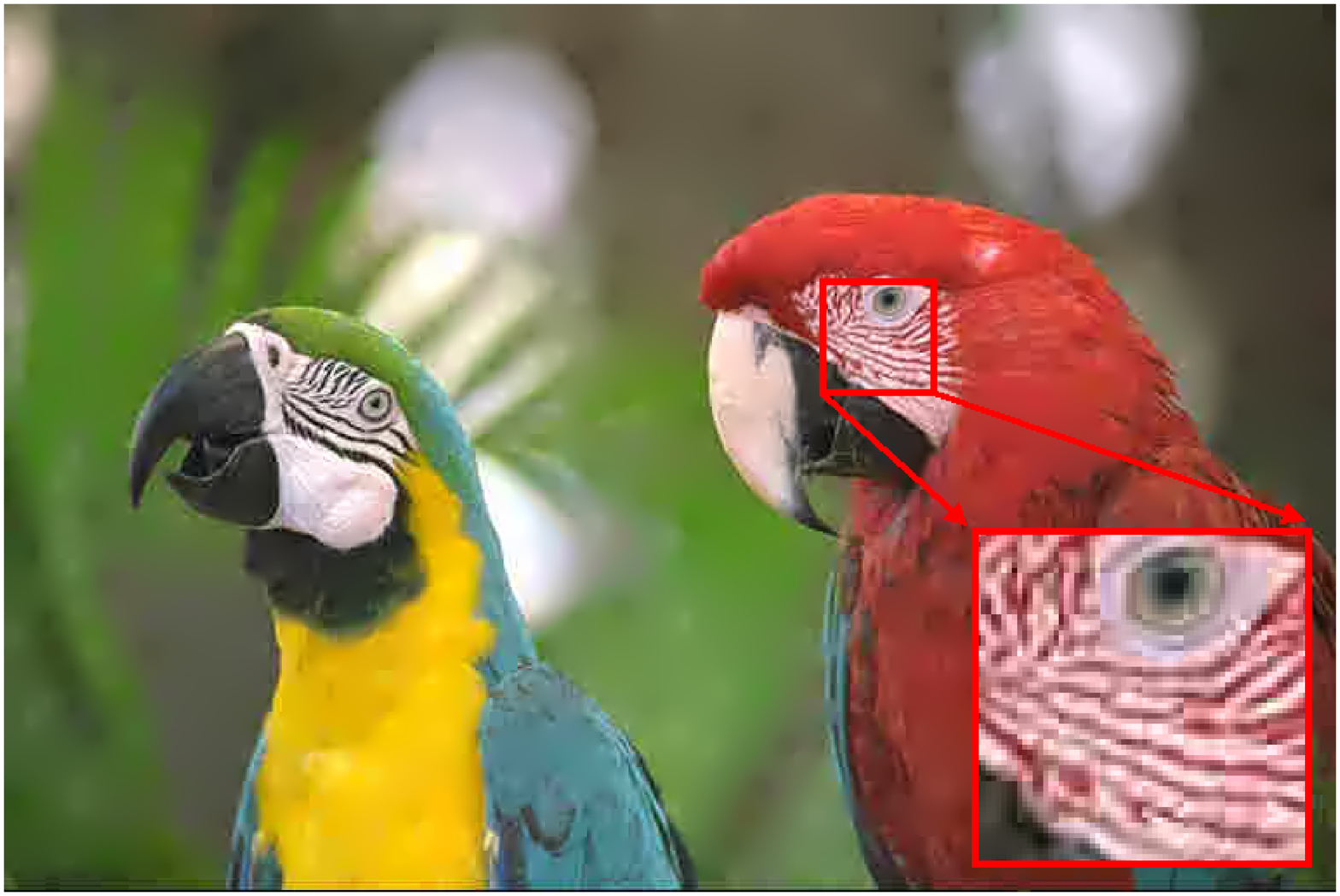} \\
		& ratio 111.1:1, PSNR 27.8 dB & ratio 111.1:1, PSNR 31.4 dB 
	\end{tabular}
	\caption{Comparison between JPEG and B-EED for the images 15, 20, and 23 of the Kodak database (from top to bottom). The zooms contain areas with mostly JPEG compressed blocks. In addition to increasing the PSNR, B-EED reduces blocking artefacts in the background and also improves quality in textured regions through increased JPEG initialisation quality.
	\label{fig:kodak}}
\end{figure*}

\subsection{Storage and Reconstruction}

The B-EED encoded image has very little storage overhead. In the header, we write the image size and EED parameters $\sigma$ and $\lambda$ separately for luma and chroma channels. Since we represent inpainting block positions in a binary mask, we can store them as simple bit sequences and compress them afterwards with arithmetic coding \cite{Ri76}. For the actual image data, we interpret the remaining blocks as smaller images which we compress with the widely used Libjpeg library \cite{LIBJPEG}. The sizes of these images determine jump positions for the decoder that we additionally write to the header. 

In the reconstruction step, we read JPEG compressed blocks from the smaller JPEG images and place them at their original positions in the image. Afterwards, we apply a global EED inpainting on the original image grid for every colour channel using the information from the JPEG blocks. For the chroma channels, we guide the inpainting with the diffusion tensor of the reconstructed luma channel as in R-EED-LP. 

\section{Experimental Evaluation}\label{sec:experiments}
Our experiments cover two main topics: Firstly, we examine which image 
structures our B-EED codec picks as optimal data on a simple test image. 
Secondly, we compare B-EED against JPEG and R-EED-LP .

\subsection{EED and Corners: A Good Match}
So far, there is no codec that combines EED inpainting with stored regions 
instead of pixels. Based on the results of Schmaltz et al.\cite{SPME14}, one 
could only conjecture that a codec using EED inpainting should pick corners as 
stored data for a reconstruction with minimal error. To check this hypothesis, 
we design a simple test image (Fig.~\ref{fig:block-orig}) with eight corner 
locations lying always within one distinctive $8 \!\times\! 8$ pixel block.
We then sparsify this image with our B-EED codec while driving the density to 
the extreme such that only $8$ blocks remain (Fig.~\ref{fig:block-init}).
Note that borders only serve better visibility and are not included in the test 
image. We can see that indeed, the blocks containing corner information are the 
ones that remain. Our codec identified these regions as optimal in terms of 
reconstruction error and reconstructs the original image faithfully 
(Fig.~\ref{fig:block-recon}). This indicates that using regions containing 
image corners as known data in EED-based compression can lead to highly 
efficient codecs.

\subsection{Comparison to Other Codecs}

In our next experiment, we illustrate the potential of B-EED on a real-world 
image. As a test case, we choose Image 23 from the Kodak database \cite{Ea99}. 
Figure \ref{fig:image-comparison} shows a JPEG compressed version and a B-EED 
result acquired with the same JPEG image as initialisation. One can see that 
our codec mostly removes blocks in the background area. This yields a much 
smoother reconstruction compared to JPEG, especially visible for large 
compression ratios. B-EED keeps highly structured areas as JPEG blocks and 
propagates information from there into the empty regions. Thus, we do not 
create any block artefacts apart from the ones present between JPEG blocks. In 
this way, we increase the compression ratio by $48\%$ while decreasing the PSNR by only $1\%$.

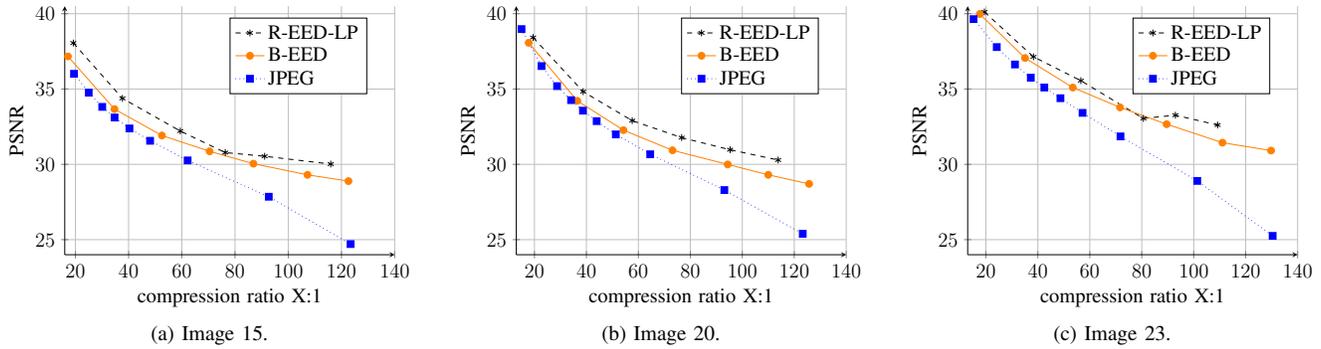
\begin{figure*}
	\centering
	\subfloat[Image 15.]{
		\scalebox{0.64}{
			\begin{tikzpicture}
		\begin{axis}
		[
		width=\axisdefaultwidth,
		height=0.8*\axisdefaultwidth,
		font=\large,
		samples=500,
		axis lines=left,
		ymin = 24, ymax = 40.5,
		xmin = 16, xmax = 140,
		ylabel={PSNR},
		xlabel={compression ratio X:1},
		xlabel near ticks,
		ylabel near ticks,
		grid = major,
		legend entries={R-EED-LP, B-EED, JPEG},
		legend cell align=left,
		legend style={at={(0.5, 0.80)}, anchor = west}
		]
		
		\addplot[dashed, black, mark=asterisk, mark options={solid, black}]
		table[x = rate, y expr=\thisrow{psnr}] 
		{figures/rd-kodim15/reed.dat};
		
		\addplot[solid, orange, mark=*, mark options={solid, orange}]
		table[x = rate, y expr=\thisrow{psnr}] 
		{figures/rd-kodim15/beed.dat};
		
		\addplot[dotted, blue, mark=square*, mark options={solid, blue}]
		table[x = rate, y expr=\thisrow{psnr}] 
		{figures/rd-kodim15/jpeg.dat};
		
		\end{axis}
		\end{tikzpicture}
	}
	}
	\hfil
	\subfloat[Image 20.]{
		\scalebox{0.64}{
		\begin{tikzpicture}
		\begin{axis}
		[
		width=\axisdefaultwidth,
		height=0.8*\axisdefaultwidth,
		font=\large,
		samples=500,
		axis lines=left,
		ymin = 24, ymax = 40.5,
		xmin = 13, xmax = 140,
		ylabel={PSNR},
		xlabel={compression ratio X:1},
		xlabel near ticks,
		ylabel near ticks,
		grid = major,
		legend entries={R-EED-LP, B-EED, JPEG},
		legend cell align=left,
		legend style={at={(0.5, 0.80)}, anchor = west}
		]
		
		\addplot[dashed, black, mark=asterisk, mark options={solid, black}]
		table[x = rate, y expr=\thisrow{psnr}] 
		{figures/rd-kodim20/reed.dat};
		
		\addplot[solid, orange, mark=*, mark options={solid, orange}]
		table[x = rate, y expr=\thisrow{psnr}] 
		{figures/rd-kodim20/beed.dat};
		
		\addplot[dotted, blue, mark=square*, mark options={solid, blue}]
		table[x = rate, y expr=\thisrow{psnr}] 
		{figures/rd-kodim20/jpeg.dat};
		
		\end{axis}
		\end{tikzpicture}
	}
	}
	\hfil
	\subfloat[Image 23.]{
		\scalebox{0.64}{
		\begin{tikzpicture}
		\begin{axis}
		[
		width=\axisdefaultwidth,
		height=0.8*\axisdefaultwidth,
		font=\large,
		samples=500,
		axis lines=left,
		ymin = 24, ymax = 40.5,
		xmin = 13, xmax = 140,
		ylabel={PSNR},
		xlabel={compression ratio X:1},
		xlabel near ticks,
		ylabel near ticks,
		grid = major,
		legend entries={R-EED-LP, B-EED, JPEG},
		legend cell align=left,
		legend style={at={(0.5, 0.80)}, anchor = west}
		]
		
		\addplot[dashed, black, mark=asterisk, mark options={solid, black}]
		table[x = rate, y expr=\thisrow{psnr}] 
		{figures/rd-kodim23/reed.dat};
		
		\addplot[solid, orange, mark=*, mark options={solid, orange}]
		table[x = rate, y expr=\thisrow{psnr}] 
		{figures/rd-kodim23/beed.dat};
		
		\addplot[dotted, blue, mark=square*, mark options={solid, blue}]
		table[x = rate, y expr=\thisrow{psnr}] 
		{figures/rd-kodim23/jpeg.dat};
		
		\end{axis}
		\end{tikzpicture}
	}
	}
	
	\caption{Rate distortion curves for three images from the Kodak database. 	   B-EED outperforms JPEG consistently and comes close to R-EED-LP with less computational effort.
		\label{fig:ratedistortion}} 
\end{figure*}

In our third experiment, we visually and quantitatively evaluate B-EED on a selection of images from the Kodak database. Along with Image 23, we select Image 20 and 
Image 15 as representatives for varying image content.
Figure \ref{fig:kodak} shows a visual comparison with JPEG for different compression ratios. By smoothly inpainting homogeneous regions, B-EED removes blocking artefacts. Since it reduces storage cost via block removal, B-EED can choose lower compression ratios for its JPEG initialisation image leading to more accurate reconstructions also in textured regions. For the presented images, the chosen JPEG initialisation compression ratios are about $25\%$ smaller than original JPEG ones. Corresponding rate distortion curves in Figure \ref{fig:ratedistortion} indeed show that B-EED outperforms JPEG over all compression ratios, for high ratios by a large margin. It can be seen as competitive to R-EED-LP while being less computationally demanding and easier to optimise, decreasing runtime by roughly a factor 10 for medium compression ratios. In the compression pipeline described in Section \ref{sec:proposed}, the optimisation of $\sigma$ and $\lambda$ improved the error by up to $3\%$ compared to the standard choice. Furthermore, NLBE decreased the error additionally by up to $10\%$. We observed stronger improvements for higher compression ratios, since non-optimally chosen blocks in the probabilistic sparsification have a higher influence on the global error for smaller mask densities.


\section{Conclusions}
\label{sec:conclusions}

We have proposed the first hybrid B-EED codec which combines JPEG and EED inpainting. Despite its simplicity, the codec is able to outperform JPEG on a set of natural images. Additionally, it performs similarly as the state of the art of PDE-based codecs while using much less demanding optimisation techniques. This shows that already simple components can lead to viable hybrid codecs performing well on a variety of image content. We expect that the application of our proposed concepts to more sophisticated codecs such as H.264~\cite{SW05} can lead to similar improvements.

Furthermore, our work is the first to show that a codec combining EED inpainting with stored regions instead of pixels automatically selects image corners as optimal known data. This opens up a number of possibilities for advanced data selection strategies for PDE-based codecs which supplement pixel information with additional corner regions.

\bibliographystyle{IEEEbib}
\bibliography{myrefs}

\end{document}